\documentclass[journal=jctcce,manuscript=article,layout=twocolumn]{achemso}
\usepackage[version=3]{mhchem} 
\usepackage{graphicx}
\usepackage{subcaption}

\makeatletter
\let\l@addto@macro\relax
\makeatother
\usepackage[fontsize=10pt]{scrextend}

\author{Jayashrita Debnath}
\affiliation[]{Department of Chemistry and Applied Biosciences, ETH Zurich, c/o USI Campus, Via Giuseppe Buffi 13, CH-6900, Lugano, Switzerland}
\alsoaffiliation{Facolt\`{a} di Informatica, Istituto di Scienze Computazionali,  Universit\`{a} della Svizzera Italiana (USI), Via Giuseppe Buffi 13, CH-6900, Lugano, Switzerland}
\author{Michele Parrinello}
\affiliation{Department of Chemistry and Applied Biosciences, ETH Zurich, c/o USI Campus, Via Giuseppe Buffi
 13, CH-6900, Lugano, Switzerland}
\alsoaffiliation{Facolt\`{a} di Informatica, Istituto di Scienze Computazionali,  Universit\`{a} della Svizzera Italiana (USI), Via Giuseppe Buffi 13, CH-6900, Lugano, Switzerland}
\alsoaffiliation{Instituto Italiano di Tecnologia, Via Morego 30, 16163 Genova, Italy}
\email{parrinello@phys.chem.ethz.ch.}

\title[An \textsf{achemso} demo]
  {Gaussian Mixture Based Enhanced Sampling}

\begin{document}



\begin{abstract}
Many processes in chemistry and physics take place on timescales that cannot be explored using standard molecular dynamics simulations. 
This renders the use of enhanced sampling mandatory. Here we introduce an enhanced sampling method that is based on constructing a model probability density from which a bias potential is derived. The model relies on the fact that in a physical system most of the configurations visited can be grouped into isolated metastable islands. To each island we associate a distribution that is fitted to a Gaussian mixture. The different distributions are linearly combined together with coefficients that are computed self consistently. Remarkably, from this biased dynamics, rates of transition between different metastable states can be straightforwardly computed.

\end{abstract}

\vspace{0.5cm}
One of the most active areas of theoretical chemical physics is that of enhanced sampling, especially in the area of atomistic simulations. The roots of this interest lie in the presence of kinetic bottlenecks in many of the systems of current interest. These bottlenecks drastically reduce the probability of observing transitions between different metastable states in an affordable computer time, thus limiting the possibility of studying interesting and important phenomena like chemical reactions, nucleation, and protein plasticity.

In order to overcome this limitation a great variety of methods has been suggested \citep{BaronPeters}. The first such method has been described in the classic work of Torrie and Valleau\citep{Torrie1977} in which umbrella sampling has been introduced. In this paper a bias potential $V({\boldsymbol{R}})$, function of the atomic coordinates ${\boldsymbol{R}}$, is added to the potential $U({\boldsymbol{R}})$. The role of $V({\boldsymbol{R}})$  is to facilitate transitions from one metastable state to another. The Boltzmann expectation value of any operator $O({\boldsymbol{R}})$ is then computed from the biased trajectories using the reweighting formula:
\begin{equation}
    \label{ensemble}
    \Big< O({\boldsymbol{R}}) \Big> = \frac{\Big< O({\boldsymbol{R}}) \ e^{\beta V({\boldsymbol{R}}) } \Big>_V}{\Big<  e^{\beta V({\boldsymbol{R}})} \Big>_V}
\end{equation}

where $\beta$ is the inverse temperature and the suffix V indicates that the averages are performed over the ensemble biased by $V({\boldsymbol{R}})$. Torrie and Valleau suggested writing $V({\boldsymbol{R}})$ in the form $V({\boldsymbol{s}({\boldsymbol{R}})})$ where $s({\boldsymbol{R}})$ is a set of order parameters or collective variables (CVs) that describe the difficult-to-sample degrees of freedom. After Torrie and Valleau many proposals have been put forward on how to build a suitable $V({\boldsymbol{s}({\boldsymbol{R}})})$. Among the many methods suggested, one can mention here adaptive umbrella sampling\citep{adaptiveUS}, Gaussian mixture umbrella sampling\citep{GMM-US}, metadynamics\citep{Laio2002} and variationally enhanced sampling\citep{Valsson2014}.  
Of course, crucial to the success of all of these methods, is the use of an appropriate set of CVs. While the process of identifying CVs can be very insightful \citep{Gervasio}, it can also be time consuming, in spite of the fact that several methods have been proposed to facilitate the CV construction\citep{Mendels2018,SMLCV,Tiwary2015a,McCarty2017a,Sultan2017}. 

Here we take a different approach that does not require the introduction of a restricted set of CVs and aims at constructing an intrinsically multidimensional bias.  We borrow here the strategy of adaptive umbrella sampling and rather than building the bias we operate first on the probability distribution and then we link it to the bias. 

\begin{equation}
    \label{bias_prob}
    V({\boldsymbol{R}})=\frac{1}{\beta} \log P_m({\boldsymbol{R}}) ,
\end{equation}
Similar strategies have been followed also in a number of publications\citep{RAVE,GMM-US,ReconnaisanceMTD}. However we differ from this class of methods in that we do not assume that the bias depends on a very restricted set of CVs. 

In order to understand why one establishes such a link, we consider two extreme cases. In the first, $P_m({\boldsymbol{R}})$ coincides with the Boltzmann distribution
\begin{equation}
    \label{boltzmann}
    P_m({\boldsymbol{R}}) = \frac{e^{-\beta U({\boldsymbol{R}})}}{Z}
\end{equation}
$Z = \int d{\boldsymbol{R}} \ e^{-\beta U({\boldsymbol{R}})} $ being the partition function. In such a case, apart from an irrelevant constant, $V({\boldsymbol{R}})=-U({\boldsymbol{R}})$. This amounts to performing a uniform sampling. This is in the practice a useless endeavour. In the second case we take $P_m({\boldsymbol{R}}) = 1$ thus $V({\boldsymbol{R}})=0$ and this amounts to imposing a null bias. Of course a useful choice is a compromise between these two extremes, too much bias and no bias.

In setting up our density model we have in mind the fact that, as discussed earlier, a physical system spends most of its time visiting only a 
finite number of metastable states. 
We characterize the metastable states by a number of descriptors $\boldsymbol{d}({\boldsymbol{R}})\equiv \{d_i({\boldsymbol{R}})\ ; i \in \{1,N_d\} \}$ that delimit the space in which we want the bias to act. 
Identifying a set of useful descriptors is far less challenging than determining a set of CVs. In fact $N_d$ can be large, while the number of CVs needs to be small since the computational cost scales exponentially with the number of CVs. In addition the choice of CVs implies a hypothesis, even if tentative, on the transition mechanism. No such insight is needed here and the  $\boldsymbol{d}({\boldsymbol{R}})$ is just a set of variables that can distinguish between the different metastable states, in such a way that equilibrium configurations belonging to different metastable states are projected into separate regions. 
Implicitely we are assuming here that the rate of transition between states is much slower than their internal dynamics.

For each of the $M$ metastable states we then run short trajectories obtaining $M$ sets of configurations. We use these configurations to model the probability density $p^i(\boldsymbol{d})$ for each metastable state $i$. Here we  approximate $p^i(\boldsymbol{d})$ by a Gaussian mixture but other choices are also possible.

In the Gaussian mixture scheme $p^i(\boldsymbol{d})$ is expressed as a linear combination of multivariate Gaussians:
\begin{equation}
    \label{Gaussianmixture}
    p^i(\boldsymbol{d}) \cong \sum\limits_{k=1}^{K^i} \pi_k^i \  \mathcal{N}(\boldsymbol{d}|\boldsymbol{\mu}_k^i,\boldsymbol{\Sigma}_k^i)
\end{equation}
where the mixing coefficients $\pi_k^i$  satisfy the conditions $0\leq \pi_k^i \leq 1$ and $\sum\limits_{k=1}^{K^i} \pi_k^i = 1 $ \citep{Bishop} . Having determined the M $p^i({\boldsymbol{d}})$, we then construct $P_m({\boldsymbol{R}})$ as
\begin{equation}
    \label{probability}
    P_m({\boldsymbol{R}}) = \frac{1}{M} \sum\limits_{i=1}^M \frac{p^i(\boldsymbol{d({\boldsymbol{R}})})}{Z^i}
\end{equation}
where we have introduced the normalizing constants $Z^i = \int d{\boldsymbol{R}}\  p^i(\boldsymbol{d}({\boldsymbol{R}}))$.


The bias then becomes:
\begin{equation}
    \label{bias}
    V({\boldsymbol{R}}) = \frac{1}{\beta} \log \Big( \frac{1}{M} \ \sum\limits_{i=1}^M \frac{p^i(\boldsymbol{d({\boldsymbol{R}})})}{Z^i}  \Big)
\end{equation}
which, after dropping irrelevant constants, can be rewritten as:
\begin{equation}
    \label{bias-f}
    V({\boldsymbol{R}}) = \frac{1}{\beta} \log \Big( \sum\limits_{i=1}^M \frac{Z^1}{Z^i} \ p^i(\boldsymbol{d({\boldsymbol{R}})})  \Big)
\end{equation}
The ratios $Z^1/Z^i$ are not known a priori, but they can be set to an arbitrary value at the beginning of calculation and estimated self consistently from
\begin{equation}
    \label{factors}
    \frac{Z^1}{Z^i} = \frac{\Big< p^1(\boldsymbol{d({\boldsymbol{R}})}) e^{\beta(U({\boldsymbol{R}})+V({\boldsymbol{R}}))}\Big>_{V}}{\Big< p^i(\boldsymbol{d({\boldsymbol{R}})}) e^{\beta(U({\boldsymbol{R}})+V({\boldsymbol{R}}))}\Big>_{V}}.
\end{equation}
In doing so we are taking inspiration from integrated tempering sampling \citep{ITS,MetaITS}. Alternatively one could use the variational method in Ref. \citenum{VESdF} or its possible generalization to many states. We note that the introduction of the terms $Z^1/Z^i$ is crucial for getting a good bias potential since it gives information on the relative free energy differences between states.  However, accurately converged values of $Z^1/Z^i$ are not needed provided that, when inserted in Eq. \ref{bias-f}, the resulting bias is able to promote transitions between metastable states. The statistics is then accumulated by using Eq. \ref{bias-f} where the $Z^1/Z^i$ ratios are fixed. We refer to our method as Gaussian Mixture Based Enhanced Sampling (GAMBES). 

As it is written in equation\eqref{bias-f}, $V({\boldsymbol{R}})$ is not useful since by construction there is negligible overlap between different $p^i(\boldsymbol{d})$ and when the system visits a region of small overlap, the bias becomes much too large. Thus we do not allow the Gaussian to go to zero but we let each Gaussian decay smoothly to a preassigned value $p_c$ (see supporting information (SI) ). Thus each $p^i(\boldsymbol{d})$ acts in a limited region of space and is constant in between.  Occasionally even this remedy is not enough and in spite of the bias the system is not capable of escaping a metastable state. In such a case, we collect for a while the configurations thus accumulated. We fit the descriptor distribution of these configurations to a new Gaussian mixture and treat formally this set of configurations as a ghost metastable state. In the SI, we exemplify how this strategy works in the practice. We stress here that in our experience this procedure is rarely needed as it is the case in the examples illustrated in the main text. 

In all circumstances, in the interstitial regions the bias has a constant value $V_0$ and differs from this value whenever visiting a metastable state. By subtracting this constant from the bias one has by construction a bias that is zero in the transition state region. Thus we are in the position to make use of the ideas of conformational flooding\citep{conformationalFlooding}, hyperdynamics\citep{hyperdynamics}, infrequent metadynamics \citep{infreqMetad} and variational flooding\citep{VarFlooding}. From the biased trajectories in which the $Z^1/Z^i$ are kept constant, the rates can be computed directly using the hyperdynamics formula of Ref. \citenum{infreqMetad} that amounts at rescaling the biased trajectory time as follows:
\begin{equation}
    \label{rescale_time}
    \tau = \int_{0}^{t} dt' e^{\beta \ [V({\boldsymbol{R}}(t')) - V_0 ]}
\end{equation}
where $t$ is the simulation time and $\tau$ is the rescaled physical time.

We now first test our method on the simple but instructive case of alanine dipeptide in vacuum. At room temperature, three different conformational states are accessible, C7eq, C5, and C7ax (see SI). The peptide converts easily from C7eq to C5 and much more rarely visits C7ax. Thus we can regard this system as composed of two metastable states, in the first ($i=1$), C7eq and C5 are both populated, while in the second ($i=2$), only C7ax is visited. We run for each state two unbiased simulations of 2 ps. 

The simulations are carried out in the NVT ensemble using a molecular dynamics (MD) timestep of 2 fs and the AMBER99-SB forcefield. The temperature is kept constant at 300K using the stochastic velocity rescaling thermostat\citep{Bussi2007}. The electrostatic and non-bonded van der Waals interactions are calculated without any cutoff, and periodic boundary conditions are not imposed. All simulations are performed using GROMACS-2018.4\citep{GROMACS} and for the biased simulations, this MD engine is patched with a modified version of the PLUMED2.0 plugin\citep{Plumed2}.

We use as descriptors the two dihedral angles $\boldsymbol{d} \equiv \{\phi, \psi\}$ and fit in the probability distribution to two different multivariate Gaussian mixture models. The number of Gaussians for each state is chosen using the Bayes Information Criterion (BIC)\citep{schwarz1978}. The optimal value for state $1$ is K$^1=5$ while for state 2 it is K$^2=2$ (see SI). 

In order to evaluate the performance of the method, we run $50$ independent $10$ ns long biased simulations with different initial configurations selected from a previous biased trajectory. In order to have a reference value, we also run $50$ independent well-tempered metadynamics simulations of length 10 ns and with a bias factor of $\gamma = 10$. 

\begin{figure}
    \centering
    \includegraphics[scale=0.5]{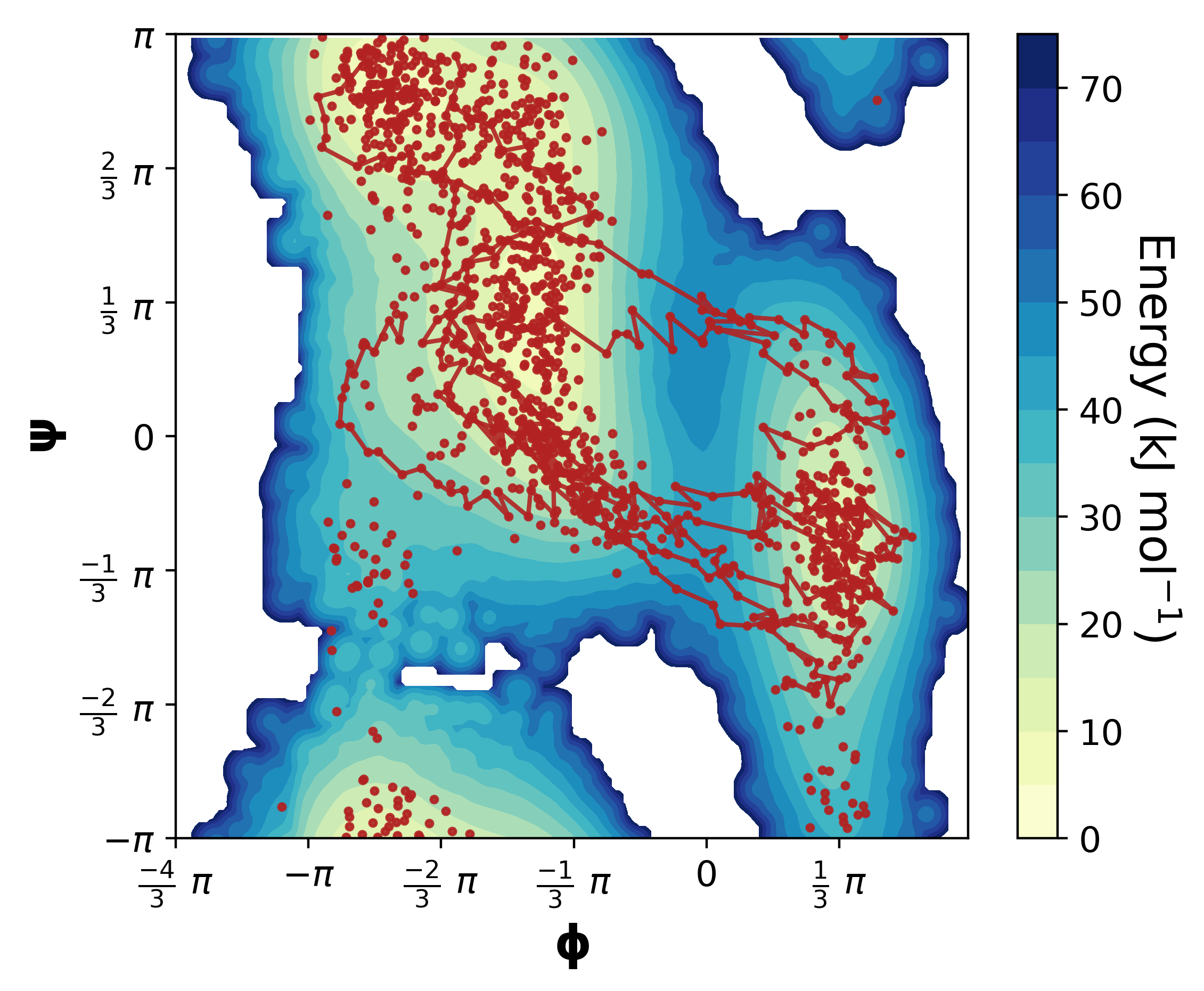}
    \caption{Free energy surface of alanine dipeptide obtained from a $10$ ns GAMBES simulation. In red are some of the points visited during the simulation. Few selected trajectories are also marked with continuous lines. }
    \label{fig:alafestrj}
\end{figure}

\begin{figure}
    \centering
    \includegraphics[scale=0.5]{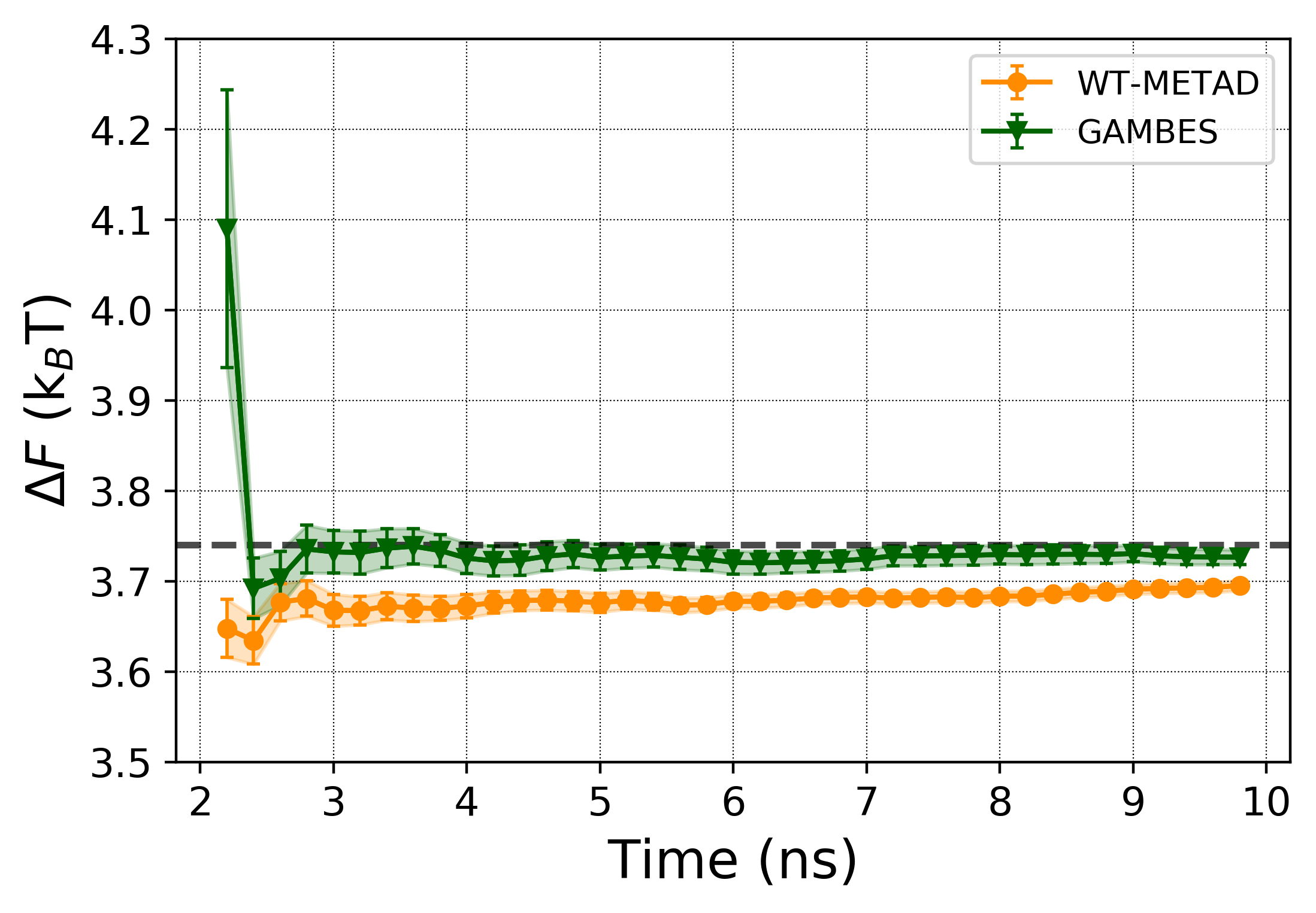}
    \caption{Free energy differences of alanine dipeptide averaged over $50$ independent simulations of GAMBES and well-tempered metadynamics (WT-METAD).  }
    \label{fig:alafree_energy_diff}
\end{figure}

In figure \ref{fig:alafestrj} we give an impression of the dynamics associated with our new method.  It can be seen that the system almost never visits the very high energy states and samples only the basins and the regions in between. Furthermore, although no directionality is imposed, the system follows the expected low free energy routes when translocating from one state to another. Additionally, out of the two possible low free energy pathways, the system prefers the lower energy route, as it should.

In figure \ref{fig:alafree_energy_diff}, the time evolution of the mean and the standard deviation of the free energy difference (see SI) between the two states are shown.
In the metadynamics runs $\Delta F$ has been computed using the reweighting scheme of \citeauthor{tiwary_reweight} \citep{tiwary_reweight}, discarding the initial $2$ ns trajectory during which the system still evolves towards the asymptotic limit in which the reweighting of Ref. \citenum{tiwary_reweight} is valid. 
In the GAMBES runs, we use the initial $2$ ns to estimate the ratio $Z^1/Z^2$ and then keep fixed this ratio for the remaining $8$ ns, during which we calculate $\Delta F$ using the static reweighting of Eq. \ref{ensemble}. It is reassuring to see that the performance of our method is even slightly better than that of metadynamics. In fact experience has shown that it is difficult to outperform metadynamics when good CVs are used. Such is the case for alanine dipeptide when both $\phi$ and $\psi$ are used as CVs. 

\begin{figure}[ht]
   \centering
    \includegraphics[scale=0.45]{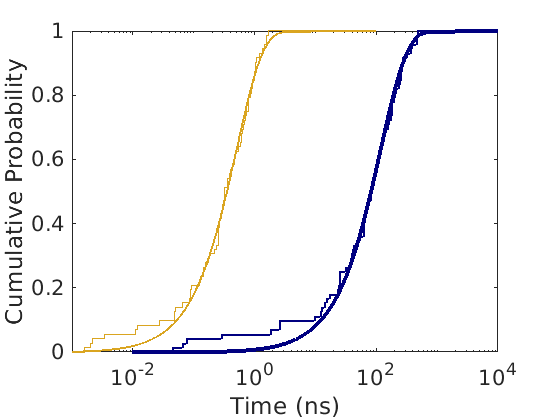}
   \caption{Cumulative probability distribution of first passage times (thin yellow) and the rescaled first passage times (thin blue) for C7eq $\rightarrow$ C7ax transition. The Poisson distribution fit for the rescaled time with $\tau_{AB}= 116$ ns is shown (thick blue). The p-value associated with the Kolmogorov-Smirnov test is found to be $0.79$}
    \label{fig:ala2_rate}
\end{figure}

We now analyze a representative trajectory in order to extract the reaction rate with the use of Eq. \ref{rescale_time}. We focus here on the rate in which the alanine dipeptide moves from the C7eq-C5 basin to the C7ax basin following the lowest free energy path. This being the transition for which most statistics can be accumulated. In figure \ref{fig:ala2_rate}, we plot the cumulative probability and extract a transition time $\tau_{AB}$ by fitting a Poisson distribution\citep{salvalaglio14,VarFlooding}. The value $\tau_{AB}=116$ ns is extracted while the values of $\tau_{AB}$ as reported in Ref. \citenum{salvalaglio14} are 110.2 and 106.3 for MD and well-tempered and infrequent Metadynamics\citep{infreqMetad} respectively. Detailed comparison of the values obtained with that of Ref.\citenum{salvalaglio14} are reported in the SI.

\begin{figure}[ht]
   \centering
    \includegraphics[scale=0.45]{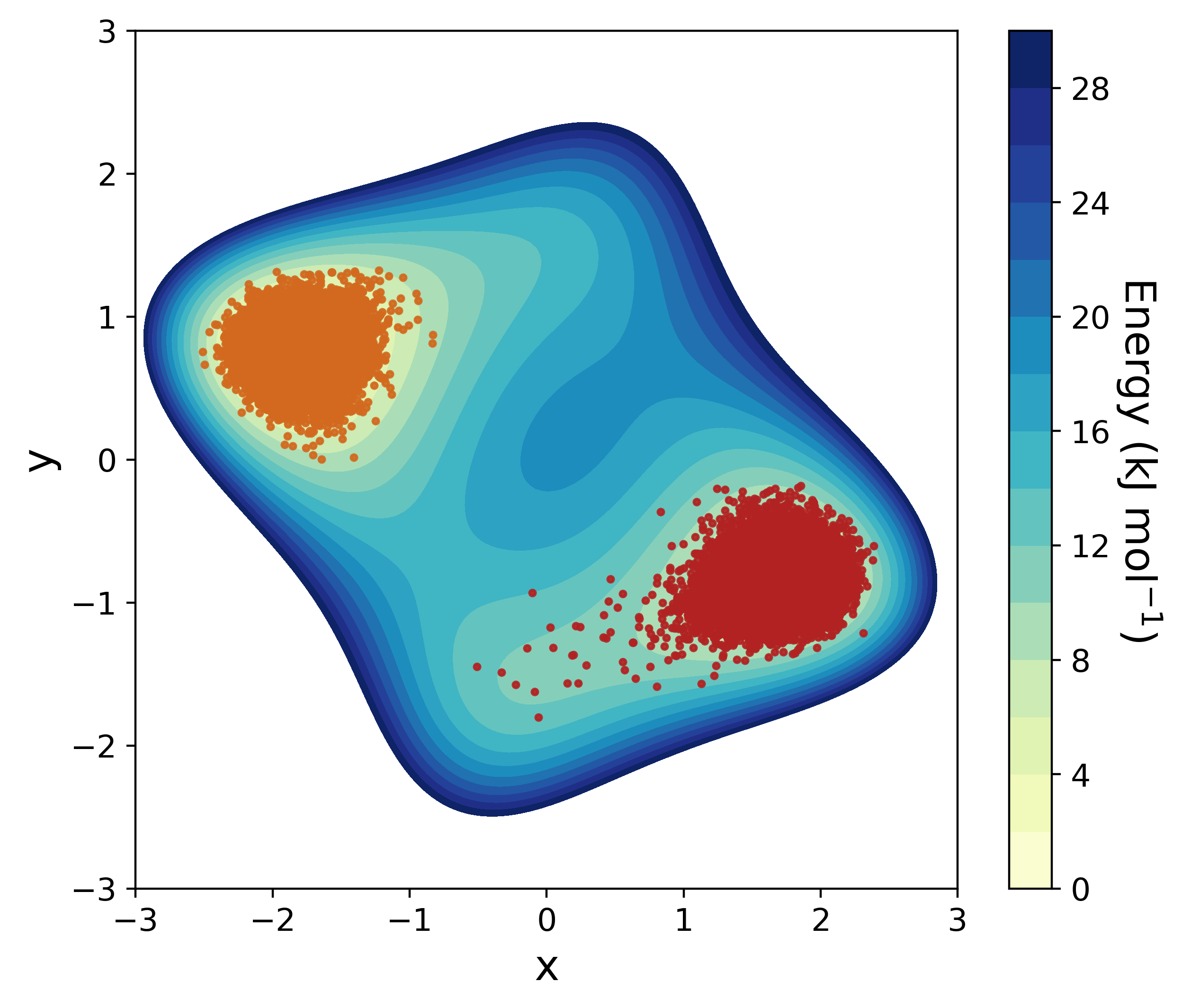}
   \caption{ Modified Wolfe-Quapp potential. The configurations visited during two unbiased trajectories started in each of the two minima are superimposed. }
    \label{fig:wolfequapp}
\end{figure}

Clearly, the ability of extracting dynamical information from ordinary biased run sets GAMBES aside from other methods. However even restricting ourselves to the consideration of its performance in the calculation of equilibrium properties, GAMBES offers considerable advantages when metadynamics deals with suboptimal CVs\citep{VESdF}. 
This being the case most encountered in the practice when dealing with real systems. For this reason we explore the performance of our method in a case in which the CVs are suboptimal. This is the case of a modified Wolfe Quapp potential\citep{VESdF} (figure \ref{fig:wolfequapp}) in which the suboptimal x coordinate is used as CV. For the sake of comparison, we ignore on purpose the fact that in GAMBES the number of descriptors can be made large at essentially zero cost and we perform a GAMBES run using only this one suboptimal CV as descriptor.
 
\begin{figure}[ht]
    \centering
    \includegraphics[scale=0.5]{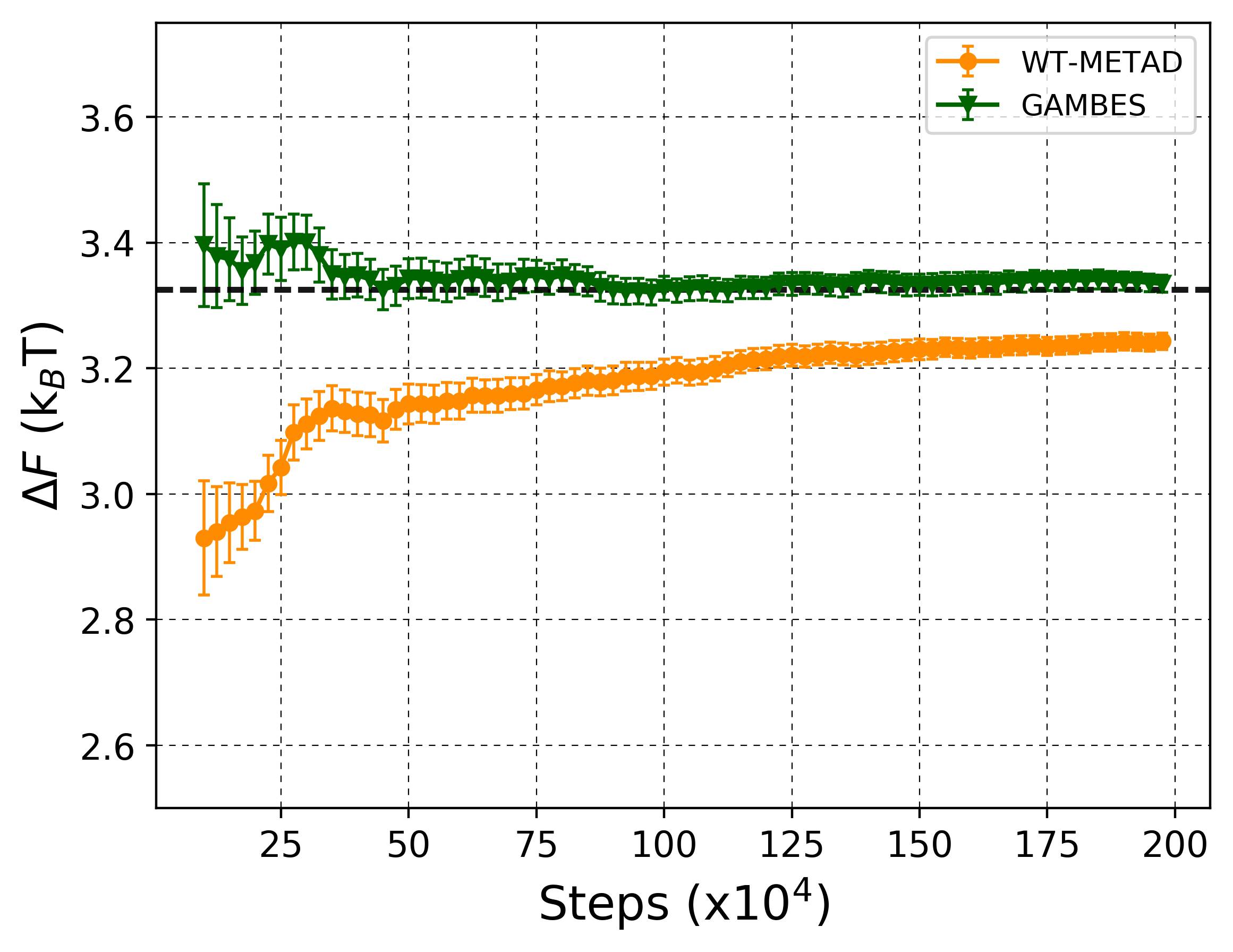}
    \caption{Time evolution of free energy differences averaged over 50 simulations. The  error bars denote the standard deviation. The black dashed line is the mean free energy difference obtained from a set of $1 \times 10^7$ steps long well-tempered metadynamics simulations.  }
    \label{fig:wolfequapp_deltaF}
\end{figure}

The calculation follows the line of the alanine dipeptide one and the technical details are given in the SI. The free energy differences averaged over the 50 simulations and their standard deviation are shown in figure \ref{fig:wolfequapp_deltaF} for the two sets of simulations. It can be seen from figure \ref{fig:wolfequapp_deltaF} that GAMBES outperforms standard well-tempered metadynamics even if its full potential is not used.
Given the suboptimal character of the CV we have not attempted to obtain reaction rates.


\begin{figure}
    \centering
    \includegraphics[scale=0.3]{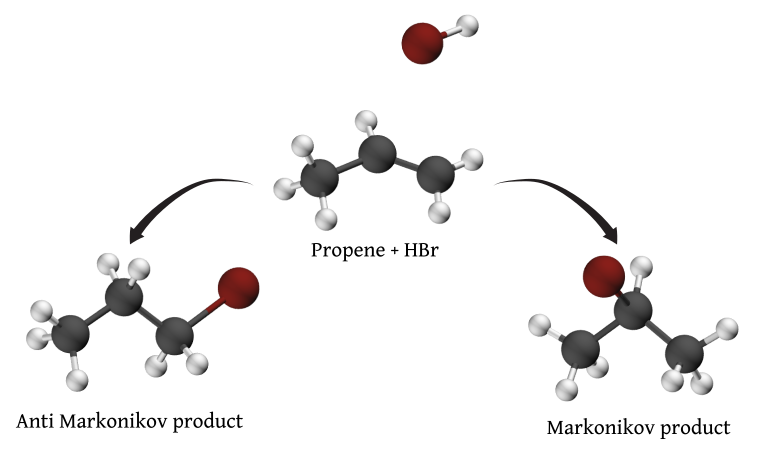}
    \caption{Schematic representation of the Hydro-bromination of propene reaction. }
    \label{fig:reaction}
\end{figure}

We now illustrate the applicability of GAMBES to a multistate system.  This is a case in which our method has some additional advantage relative to methods like Harmonic Linear Discriminant Analysis (HLDA). In fact in HLDA the number of CVs increases with the number of metastable states. 
In contrast in GAMBES the number of states can be increased without having to pay an exorbitant cost.
As a representative of the multistate case we take the hydrobromination reaction of propene (figure \ref{fig:reaction}). In this reaction,  the addition of bromine to propene can lead to two possible products due to the asymmetry of the propene carbon centers. These two products are commonly referred to as Markonikov and anti-Markonikov.

As it has been shown in Ref. \citenum{Piccini18}, studying even a simple chemical reaction like this requires at least 5 descriptors that can distinguish between the three states of interest. Following this work, we choose similar distance based decsriptors. We provide all the computational details in the SI.  

\begin{figure}
    \centering
    \includegraphics[scale=0.5]{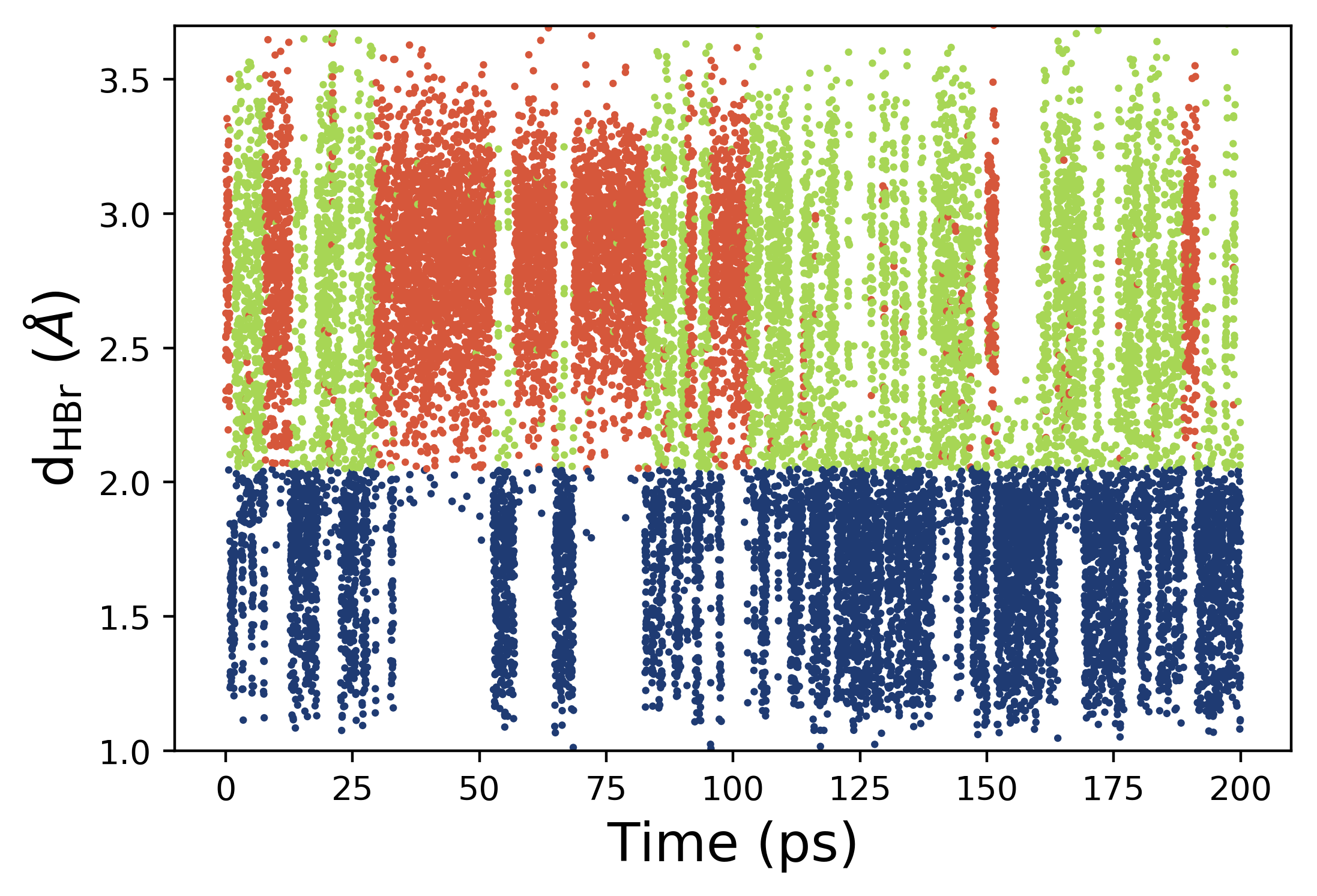}
    \caption{Time evolution of a representative descriptor (H-Br distance) as a function of the simulation time during a 200 ps long equilibration simulation. The colours blue, green and orange denote the reactant, Markonikov and the anti-Markonikov states respectively. }
    \label{fig:propbr-cv}
\end{figure}

In figure \ref{fig:propbr-cv}, we can see that in about 200 ps there have been at least 4 transitions between each state and the estimates of $Z^1/Z^2$ and $Z^1/Z^3$ are capable of inducing frequent transitions. Thus we fix the $Z^1/Z^2$ and $Z^1/Z^3$ values and run 200 ps long fixed bias simulation. 

\begin{figure}
    \centering
    \includegraphics[scale=0.5]{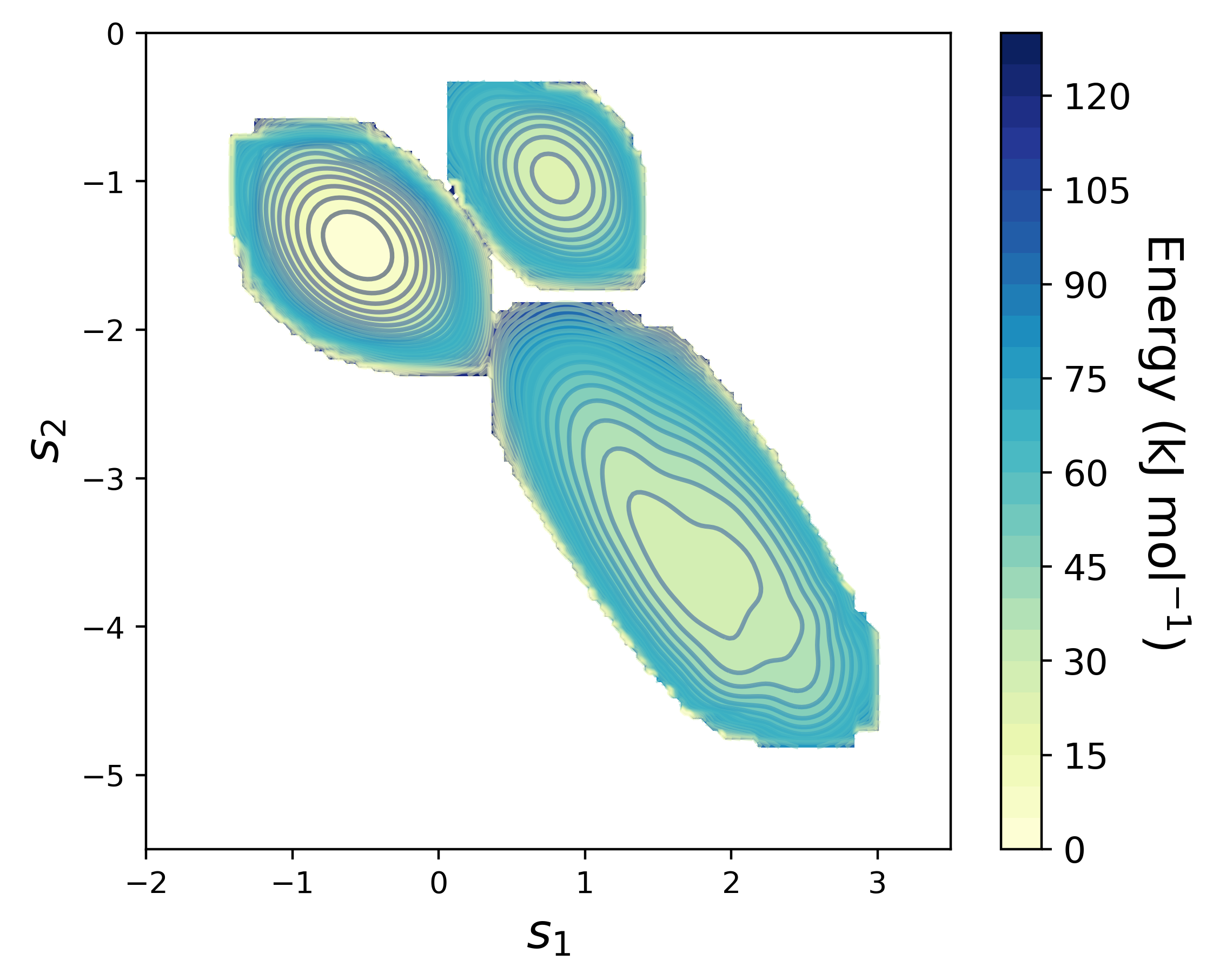}
    \caption{The free energy surface of Hydrobromination reaction obtained using GAMBES, projected on the HLDA CVs}
    \label{fig:propfes1}
\end{figure}

In order to represent the result we calculate the free energy surface along the $s_1$ and $s_2$ CVs that are obtained using the HLDA procedure (see SI). The free energy obtained using a static reweighting (Eq. \eqref{ensemble}) is shown in figure \ref{fig:propfes1}. 
For a study of the reaction rates, we refer the reader to the SI where it is shown that from the biased run, the reaction rates toward the Markonikov product can be calculated. On fitting a Poisson distribution to the rescaled first passage time, a $\tau=2.2 \times 10^7 s$  with a p-value associated with the Kolmogorov-Smirnov test of 0.92 were obtained.

In conclusion we have presented a sampling method that belongs to the family of umbrella sampling and it is based on modelling the probability distribution rather than the bias itself, making it closer in spirit to adaptive umbrella sampling. Besides providing an alternative point of view on the sampling problem, the method appears to offer distinctive advantages.  
One can use a large number of descriptors. 
Different states can be modeled with different descriptors (see SI). One could also use the method in an exploratory fashion. That is one starts with an initial assumption on the possible metastable states and then as new states are discovered their Gaussian mixture model can be added to $P_m(\boldsymbol{R})$. An area in which this procedure appears to hold great promise is that of the study of multi-step chemical reactions. Possibly the most exciting prospective is to calculate in a single shot both the static and dynamical properties.


%
\begin{acknowledgement}
This research was supported by the European  Union  Grant  No. ERC-2014-ADG-670227. Calculations were performed using the Euler HPC Cluster at ETH Z\"urich. The authors thank Michele Invernizzi for useful discussions and for providing the modified Wolfe-quapp potential. The authors thank Luigi Bonati and Michele Invernizzi for carefully reading the manuscript.
\end{acknowledgement}

{\footnotesize\bibliography{gambes-arxiv2}}

\end{document}